\documentclass{article}%
\usepackage{amsmath}
\usepackage{amsfonts}
\usepackage{amssymb}
\usepackage{graphicx}%
\providecommand{\U}[1]{\protect\rule{.1in}{.1in}}

\begin{document}

\title{On a recent very simple generalization of DMT theory of adhesion}
\author{M.Ciavarella\\Politecnico di BARI. \\V.le Gentile 182, 70125 Bari-Italy. \\Email mciava@poliba.it}
\maketitle

\begin{abstract}
As the interest in adhesion is shifting towards smaller and smaller scales,
the well known Tabor adhesion parameter decreases and the DMT theory is
frequently considered to be the appropriate limit. A very attractive much
simplified version of DMT has been proposed in a recent investigation of rough
contacts by Pastewka \& Robbins (PNAS, 111(9), 3298-3303, 2014) which seems to
work in very general conditions in numerical experiments. However, when
comparing this calculation to the known theories for the sphere, surprisingly
large conflicts occur, and the reason for the success of the numerical
experiments is obscure. 

\end{abstract}

Keywords: Adhesion, Greenwood-Williamson's theory, rough surfaces

\bigskip

\section{Introduction}

Pastewka \& Robbins (2014, PR in the following) have made a quite interesting
generalization of the DMT theory in their numerical simulations involving
atomistics rough solids. They postulate that the attractive forces have little
effect on the detailed morphology of the repulsive contact area, the repulsive
force and mean pressure are also nearly unchanged, so that they are close to
the Derjaguin-Muller-Toporov (DMT) limit (Derjaguin et al., 1975, Muller et
al., 1980, 1983). They then notice that in the "attractive" regions of area
$A_{att}$, the pressure is simply the theoretical strength of the material,
$\sigma_{th}=w/\Delta r$, where $w$ is surface energy, and $\Delta r$ is a
range of attraction. This leads them to suggest a very simple model for
adhesion, by considering the first order expansion of the separation distance
between two contacting bodies under repulsive forces only, which scales as
distance$^{3/2}$ virtually for any geometry, and equating the peak separation
to the characteristic distance $\Delta r$, they find the size of the
attractive regions. The special behaviour of rough contact leads to a very
simple picture in which characteristic sizes of repulsive and attractive
regions are defined by two almost constant length scales for any given system,
$d_{rep}$ and $d_{att}$ (their size is said to vary less than 25\% varying the
load) so that $A_{rep}=Pd_{rep}/\pi$ and $A_{att}=Pd_{att}$, where a perimeter
$P$ increases with load. The model is so simple that it permits them to
extrapolate criteria for "stickiness" in very general terms, and it is worth
comparing it with the classical DMT theory, to see the order of approximation
involved in the very simple calculation, in an instructive case.\ 

In fact, there has been so much discussion in the literature about the DMT
theory (see next paragraph), that it is surprising that PR propose such a
back-of-the envelope calculation without encountering any significant error.

\section{The DMT theories}

It is often believed that DMT is the limit for Tabor parameter
\begin{equation}
\mu=\left(  \frac{Rw^{2}}{E^{\ast2}\Delta r^{3}}\right)  ^{1/3}=\left(
\frac{Rl_{a}^{2}}{\Delta r^{3}}\right)  ^{1/3}\rightarrow0\label{tabor}%
\end{equation}
where $R$ is the radius of the sphere, $w$ is surface energy and $E^{\ast}$
plane strain modulus of the material ($l_{a}=w/E^{\ast}$ is an alternative way
to measure adhesion as a length scale).

Greenwood (2007) has clearly demonstrated that this typical statement is
\textit{"completely unwarranted: the Tabor parameter governs the transition
from rigid body behaviour to JKR behaviour"}. The reason is that the
assumptions in DMT theory make the predictions consistently worse than
assuming the sphere to be rigid, and indeed Greenwood proceeds to suggest a
"semi-rigid" theory of adhesion, in which the sphere is first assumed rigid,
the surface forces are estimated from the undeformed shape, and with these
surface forces, the displacement on the axis of the sphere is computed.

The story of these theories starts with Bradley (1932) who obtained the
adhesive force between two rigid spheres, equal to $2\pi Rw$, and Derjaguin
(1934) who reobtained this results in a simpler way with the so called
\textit{Derjaguin approximation} (the force between elements of curved,
inclined surfaces is the same as that between elements of plane, parallel
surfaces), and finally that the attractive force between spheres separated by
a gap $h_{0}$ is insensitive to the exact shape of the force between two plane
surfaces at distance $h$, i.e. $\sigma\left(  h\right)  $, as it can be
computed from
\begin{equation}
P=2\pi R\int_{h_{0}}^{\infty}\sigma\left(  h\right)  dh
\end{equation}

However, Derjaguin (1934) in the case of elastic spheres found 1/2 the
pull-off force. JKR (Johnson Kendall and Roberts 1971) developed their
different theory assuming adhesive forces occur entirely within the contact
area, obtaining $3/4$ of the Bradley pull-off value, and hence the factor
seemed to cause a conflict. Derjaguin then developed with his colleagues the
DMT\ theory (Derjaguin \textit{et al.}, 1975) which is a "thermodynamic
method" which caused few troubles. Indeed, DMT leads to Bradley result for the
force, but the attraction forces outside the contact add a contribution
("overload", according to Maugis (2000) denomination) which decreases from
$2\pi Rw$ to $\pi Rw$ when the approach increases, which was shown to be an
error intrinsic in the "thermodynamic method" by Pashley (1984). Indeed,
Pashley (1984) showed that the force between a \textit{truncated rigid} sphere
and a flat is constant and is given by $2\pi Rw$ independent on the contact
radius. Since the Hertzian profile is for any contact radius size, closer to
the flat surface than the rigid spherical profile, the force should become
larger than $2\pi Rw$ as the area of contact increases. He then integrated the
full Lennard-Jones potential as%
\begin{equation}
P=2\pi\frac{8w}{3a_{0}}\int_{a}^{\infty}\left[  \left(  \frac{a_{0}}%
{h}\right)  ^{3}-\left(  \frac{a_{0}}{h}\right)  ^{9}\right]  rdr
\end{equation}
where for the Hertzian profile at a given value of contact area $a$
\begin{equation}
h\left(  \varepsilon,a,r\right)  =\varepsilon+\frac{1}{\pi R}\left[
a\sqrt{r^{2}-a^{2}}-\sqrt{2a^{2}-r^{2}}\arctan\sqrt{\frac{r^{2}}{a^{2}}%
-1}\right]  \label{hertz}%
\end{equation}
where $\varepsilon$ is added as there is some equilibrium distance when
considering repulsive forces. Pashley then obtained that the overload in fact
increases with $a/R$, being about double, $4\pi Rw$, as $a/R=0.1$ (which is
clearly already quite high deformation for an elastic theory). Muller,
Derjaguin and Toporov (MDT, 1983) revisiting the DMT theory, used a full
Lennard-Jones potential as had been used in an earlier full numerical analysis
(MYD, Muller et al., 1980), but again using the `thermodynamic' method as in
the original DMT paper: pull-off is found when the contact area is zero only
for $\mu<0.24$, but for $\mu>0.24$ the overload increases with approach so
that the the pull-off force exceeds the Bradley value for small positive value
of the approach.

Finally, Maugis (1991, 2000) revisited the DMT theory, and suggested the
overload should not depend on approach and indeed be constant, given by the
pull-off load, $2\pi Rw$. This is not what Pashley (1984) obtains with the
force method even in the most refined form using the full L-J potential. As we
are in fact examining the generalization of DMT theory with the force method
suggested by PR, it is important to see the consequences. Greenwood (2007)
indeed suggests Maugis' version of DMT is merely an approximation, referred to
as DMT-M, and it is the \ most frequently quoted as DMT's contribution.
However, since Maugis (1991, 2000) clearly shows that DMT-M is obtained in the
limit of his Dugdale model when the Tabor parameter decreases to $0$, it is
clear that DMT-M remains the best approximation, as the force method and the
DMT approximation has some subtle problems.  Greenwood (2007) clarifies that
for small values of the approach, the force evaluation needs care, as we need
integrating the difference between the surface forces for the gap with a rigid
sphere and those for the Hertzian gap, finds the DMT model to be constantly
worse than assuming the sphere to be rigid, but does not discuss the issue of
overload further: should a better approximation remain that of the
\textit{truncated rigid} sphere and hence a constant overload?

Maugis suggests another problem in DMT theory: that continuity of stresses is
not possible if we move from the Hertzian pressure distribution, which is zero
at the edge where gap is just the equilibrium distance $\varepsilon$, to a
positive value of tension just outside: the problem doesn't arise with L-J
potential of course. He suggests the DMT limit is seen only when the region
outside the contact is strictly infinitely large, although the integral
remains finite, to $2\pi Rw$.\ More likely, Maugis in fact obtains the\ DMT
limit as the limit of his Dugdale model, rather than by integration of the
forces in the DMT sense, as otherwise he would have incurred in the troubles
described in details by Greenwood (2007). 

When making further approximations in the force method, the results cannot
certainly improve.

\section{PR - DMT model}

Returning therefore to the PR model, the simple estimate they suggest would
work extremely well for a rigid sphere. In fact, the shape of the separation
function is simply $h\left(  x\right)  =x^{2}/\left(  2R\right)  $ as there is
no deformation, and imposing the separation is equal to $\Delta r$
gives\footnote{We avoid the notation $d_{att}$ as $d_{att}$ will have to do
with an annulus shape, whereas $x_{att}$ here is a radius of a circular
region.} $x_{att}^{2}=2R\Delta r$. If we then assume the mean pressure in the
attractive region is simply $w/\Delta r$, the force $N=\frac{w}{\Delta r}\pi
x_{att}^{2}=$ $\frac{w}{\Delta r}\pi2R\Delta r=2\pi Rw$ the well known Bradley
result, independent on the estimate of $\Delta r$ in fact.

However, as we consider the case with some actual compression, the situation
is obviously more problematic. They suggest to use the first order term in the
asymptotic expansion of the separation function, which is universally given by
the same function for simple geometries like spheres, cones, or cylinders.
Indeed, the only length scale that enters is the radius of the contact area.
As the contacting region in their numerical simulations resembles one with a
2D contact with constant average diameter $d_{rep}$ , they use the standard
prefactor for a cylinder and hence use
\begin{equation}
2\frac{h\left(  x\right)  }{d_{rep}}=\frac{\sqrt{8}}{3}h^{\prime}\left(
\frac{2x}{d_{rep}}\right)  ^{3/2}\label{PR-lateral}%
\end{equation}
where $h^{\prime}$ is the slope at the contact edge, which in their theory
they estimate from a random process. In order to compare with some known
results, we shall make the comparison of this procedure in the case of a
sphere, where $h^{\prime}=d_{rep}/\left(  2R\right)  $. The first order
expansion of (\ref{hertz}), $h\simeq\frac{\sqrt{2a}}{R}\frac{8}{3\pi}x^{3/2}$,
differs indeed only by the prefactor $\frac{8}{3\pi}=\allowbreak0.85$ being a
little larger than PR equation (\ref{PR-lateral}) considering $a=d_{rep}/2,$
which gives a prefactor $\frac{2}{3}$ (we have omitted the constant term
$a_{0}$ which in a DMT model with no repulsive force should not be included).
We shall maintain the prefactor in the original PR equation, as they suggest
it is more useful for rough surface typical asperity contacts. PR discuss that
$\Delta r$ can be estimated for arbitrary potential,\ but is of the order of
$a_{0}$, or a little smaller. With the 9-3 Lennard-Jones potential $\Delta
r=1.15a_{0},$ while for 12-6 LJ, $\Delta r=0.62a_{0}$. If we equate this
separation law with the range of attractive forces, $\Delta r$, we obtain the
lateral distance defining the size of attractive region (which is a circular
annulus of size $d_{att}$)%
\begin{equation}
d_{att}=\left[  \frac{\left(  \frac{3}{2}R\Delta r\right)  ^{2}}{d_{rep}%
}\right]  ^{1/3}%
\end{equation}
It is clear that this equation, having been obtained with a first order
expansion in $h\left(  x\right)  $, is valid if $d_{att}<<d_{rep}$, and this
introduces obviously a strong limitation when the contact radius is small, and
also a maximum radius of the sphere for any given $d_{rep}$: for
$d_{rep}=2\Delta r=2a_{0}$ , we need $\frac{R}{a_{0}}<2.66$; for
$d_{rep}=200a_{0},$ then $\frac{R}{a_{0}}<26500$. For large contact or large
compression, a stricter limit on radius comes from the range of small Tabor
parameters. For example, for $l_{a}/a_{0}=0.05$ of the LJ potential, the
assumption $\mu<1$ in (\ref{tabor}) implies $\left(  \frac{R}{a_{0}}\right)
<400$.

If we now consider the spherical case, following exactly the steps in PR
theory, the attractive load is%
\begin{equation}
P_{att}=\pi d_{rep}d_{att}\frac{w}{\Delta r}=\pi d_{rep}\frac{w}{\Delta
r}\left[  \frac{\left(  \frac{3}{2}R\Delta r\right)  ^{2}}{d_{rep}}\right]
^{1/3}=3^{2/3}\pi wR\left(  \frac{\delta}{\Delta r}\right)  ^{1/3}\label{PR1}%
\end{equation}
where $\delta$ is the compression (or approach) of the sphere, given by Hertz
theory as $\delta=d_{rep}/\left(  2R\right)  $. This model is different from
the DMT-M model for a sphere with constant overload. The simplifications we
have adopted in the analysis are evident at small compressions, as there is no
adhesive force for zero compression, unlike all the DMT theories (which of
course consider a better integration of the force) and pull-off of the
classical DMT theory is obtained only (unrealistically) for positive
compression
\[
\frac{\delta^{\ast}}{\Delta r}=\frac{8}{9}%
\]

Above this point, with further compression, the adhesive overload continues to
increase, but this is not due to the approximation of taking the first order
in the separation function. For example, for $d_{rep}=200a_{0}$, assuming the
$\mu=1$ case $R=400a_{0},$
\begin{equation}
P_{att}=2\pi wR\frac{3^{2/3}}{2}\left(  \frac{200^{2}}{4\ast400}\right)
^{1/3}=2\pi wR\times3.04
\end{equation}
and increase of a factor 3. Notice that in this case $\frac{d_{rep}/2}%
{R}=\frac{100}{400}=0.25$ and hence extrapolating the results of Fig.3 in
Pashley (1984) obtained with the full 3-9 L-J potential and the full
expression of Hertzian separation function, this result seems extremely
similar: this is another source of error, intrinsic in the force method. 

\section{Dimensionless notation}

We can illustrate the PR theory for the sphere more conveniently with
dimensionless notation using Tabor parameter%
\begin{equation}
\widehat{\delta}=\delta/\left(  \mu\Delta r\right)  \quad;\quad\widehat
{P}=P/\left(  \pi Rw\right)
\end{equation}
and (\ref{PR1}) becomes%
\begin{equation}
\widehat{P}_{att}=3^{2/3}\left(  \widehat{\delta}\mu\right)  ^{1/3}%
\end{equation}
The point where there is exact coincidence with DMT-M is now $\widehat{\delta
}=\left(  \frac{2}{3^{2/3}}\right)  ^{3}/\mu=0.89/\mu$ which obviously
increases without limit for low $\mu$.

Adding the Hertzian contribution, and writing also DMT-M (constant overload),
and the JKR theory (Johnson, et al., 1971)\footnote{JKR is presented in a
curve fitted form in order to be easily used.}
\begin{align}
\widehat{P}_{DMT-M} &  =\frac{4}{3\pi}\widehat{\delta}^{3/2}-2\\
\widehat{P}_{PR} &  =\frac{4}{3\pi}\widehat{\delta}^{3/2}-3^{2/3}\left(
\widehat{\delta}\mu\right)  ^{1/3}\\
\widehat{P}_{JKR} &  =\widehat{P}_{0}-1.1\left(  \widehat{\delta}%
-\widehat{\delta}_{0}\right)  ^{1/2}+0.43\left(  \widehat{\delta}%
-\widehat{\delta}_{0}\right)  ^{3/2}%
\end{align}
where $\widehat{\delta}_{0}=-\frac{3}{4}\pi^{2/3},\widehat{P}_{0}=5/6$ are the
JKR values at pull-off in displacement control.

The plots are shown in Fig.1 for $\mu=0.1,0.5,1$. The error for small
approaches was expected, since the first order expansion of the separation gap
tends to work less well there, and one should integrate carefully the
forces.\ Another difference reducing the adhesion forces comes from the
prefactor in the assumed form of separation, which reduces the size of
attractive regions. However, reduction of adhesive forces does not come only
for the prefactor, as it is clear from the equations above: the DMT-PR theory
simply for Tabor parameter near zero leads to the Hertz theory $\left(
\widehat{\delta}\mu\right)  =0$, which is here not plotted for clarity of
representation, as it is simply the DMT-M curve shifted up by a constant
factor 2. Further, the overload increase becomes quite large for intermediate
Tabor parameters, and conflicts with the most established form of DMT theory
(that given by Maugis).

Hence, there is only a relatively narrow range of regimes where this
approximated theory can give reasonable results. 

\begin{center}
$%
\begin{array}
[c]{cc}%
{\includegraphics[
height=1.9206in,
width=3.6601in
]%
{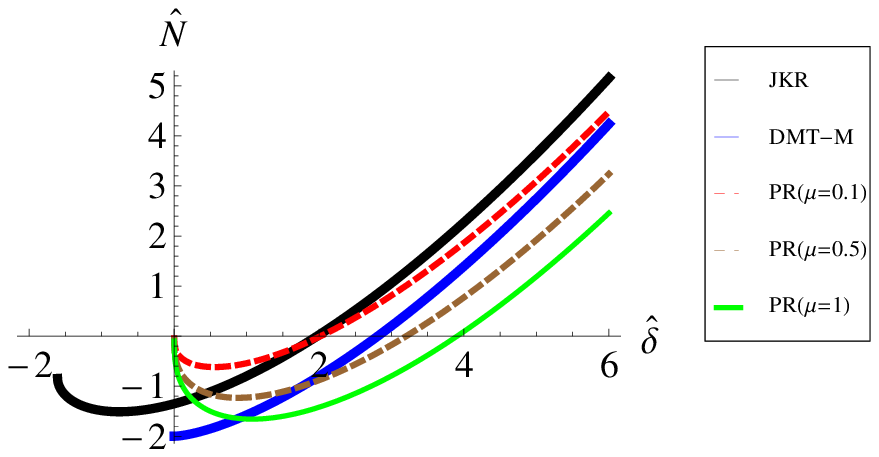}%
}
&
\end{array}
$

Fig.1. Curves load-approach for JKR, DMT-M, and DMT-PR theories, for the
sphere. PR is plotted for Tabor parameter $\mu=0.1,0.5,1$
\end{center}

\section{Discussion}

Given the extensive simulations in the PR paper, it is not clear how the PR
version of the DMT model can work reasonably well over different conditions,
including a realistic range of $\mu$:- PR simulations are said to be accurate
for Maugis version of Tabor parameter $\lambda=\left(  \frac{9}{2\pi}\right)
^{1/3}\mu=1.13\mu$ which exceeds unity (they do not say unfortunately what is
the inferior limit). It is true that DMT theories (or "force methods", or
perhaps Derjaguin--Pashley theories as Greenwood (2007) suggests)\ have not
been discussed in details other than in the pull-off limit value, but the
Maugis solution is clearly shown to tend to the DMT-M limit at small $\mu$,
and intermediate solutions with the Maugis model are indeed intermediate
between DMT-M and JKR. Here, the approximations induced by assuming that no
tension is present in the contact region, that the separation is only due to
compressive forces, and that the tension is equal to the average value
$w/\Delta r$ in the attractive regions, estimated by the first order expansion
of displacements, seem to lead to quite conflicting results with respect to
the established theories in the entire range of $\mu=0-1$ which we are
discussing. Given the DMT-PR theory is then compared with direct simulations
using the adhesive potentials, it is not clear how a conflict cannot emerge in
numerical simulations, other than perhaps a fortuitous series of error
cancellations, or by looking at special range of parameters. There is a high
risk however in using this back-of-the-envelope kind of calculation in general conditions.

Obviously, PR geometry of each individual contact probably differs in some
cases from that of a sphere, and indeed they suggest they have geometries of
contact areas closer to elongated or even "fractal" shapes: whether this can
explain the discrepancies in general is not obvious to say. From the typical
case of Fig.1, they are looking at nominal area ratios of about 2\%, and their
bandwidth can be estimated, depending on the system they are studying, to vary
between $\alpha=16$ and $1660$. For the low bandwidth cases, the asperity
models describe accurately the geometry in contact at these area ratios
(Carbone \&\ Bottiglione, 2008), and therefore the system should be very close
to an ensemble of Hertzian asperities. It is unclear how therefore their
results can be still independent over these wide range of configurations, on
Tabor parameter, and on the compression of asperities. Their potentials are
truncated in the simulations, but results with full Lennard-Jones potential do
not show significant differences, even this factor seems to be ruled out.
Interaction effects due to remote forces are possibly causing some different
relation between local force and local approach, but this would not change the results.

Notice that for Tabor parameter $\mu=1$, the x scales in Fig.1 is simply the
number of atomic steps of compression, and therefore it is meaningless to look
at $\widehat{\delta}=\delta/\left(  \mu\Delta r\right)  <1$. Hence, the
adhesive force should be generally\textit{ larger} than real. For smaller
Tabor parameters $\widehat{\delta}=\delta/\left(  10\Delta r\right)  =1$ equal
to 1 the scale in the plot is a multiple of the number of atomic steps and
hence it is realistic to say that the adhesive force will generally be
\textit{smaller} than real. Could these two errors balance somehow?

\section{Conclusion}

We have considered a recent proposal of a very simple extension of the DMT
model to general contact problems, including rough contacts, by Pastewka and
Robbins. We find, in the known case of sphere, very large possible errors, in
part already found in the literature and due to the approximations in DMT
theory, which are here exacerbated because of the further simplifications.
There is a danger therefore in using these ideas in general.

\section{References}

R.S. Bradley, (1932) The cohesive force between solid surfaces and the surface
energy of solids. Phil. Mag. 13, 853.

G.Carbone, \& F.Bottiglione, (2008). Asperity contact theories: Do they
predict linearity between contact area and load?. Journal of the Mechanics and
Physics of Solids, 56(8), 2555-2572.

B.V. Derjaguin, (1934) Theorie des Anhaftens kleiner Teilchen. Kolloid
Zeitschrift 69, 155.

\bigskip\bigskip B.V. Derjaguin, V.M. Muller and Yu.P. Toporov, (1975) Effect
of contact deformations on the adhesion of particles. J. Colloid Interface
Sci. 53, 314.

JA. Greenwood, (2007). On the DMT theory. Tribology Letters, 26(3), 203-211.

D. Tabor, (1977) Surface forces and surface interactions. J. Colloid Interface
Sci. 58, 2.

D. Maugis and M. Barquins, (1978). Fracture mechanics and the adherence of
viscoelastic bodies. J. Phys. D (Appl. Phys.) 11,  1989.

V.M. Muller, V.S. Yuschenko and B.V. Derjaguin, (1980). On the influence of
molecular forces on the deformation of an elastic sphere and its sticking to a
rigid plane. J. Colloid Interface Sci. 77, 91.

V.M. Muller, B.V. Derjaguin and Yu.P. Toporov, (1983) On two methods of
calculation of the force of sticking of an elastic sphere to a rigid plane.
Colloids Surf 7. 251.

M.D. Pashley, (1984). Further consideration of the DMT model for elastic
contact. Colloids Surf 12,  69.

KL Johnson, K. Kendall, and A. D. Roberts. (1971). Surface energy and the
contact of elastic solids. Proc Royal Soc London A: 324. 1558.

D Maugis,  (2000). Contact, adhesion and rupture of elastic solids (Vol. 130).
Springer, New York.

M. D.Pashley,  (1984). Further consideration of the DMT model for elastic
contact. Colloids and surfaces, 12, 69-77.

L. Pastewka, \& M.O.\ Robbins, (2014). Contact between rough surfaces and a
criterion for macroscopic adhesion. Proceedings of the National Academy of
Sciences, 111(9), 3298-3303.

\end{document}